# A frequentist two-sample test based on Bayesian model selection

**Pietro Berkes and József Fiser**
Volen Center for Complex Systems
Brandeis University, Waltham, MA 02454

## Abstract

Despite their importance in supporting experimental conclusions, standard statistical tests are often inadequate for research areas, like the life sciences, where the typical sample size is small and the test assumptions difficult to verify. In such conditions, standard tests tend to be overly conservative, and fail thus to detect significant effects in the data. Here we define a novel statistical test for the two-sample problem. Several characteristics make it an attractive alternative to classical two-sample tests: 1) It is based on Bayesian model selection, and thus takes into account uncertainty about the model's parameters, mitigating the problem of small samples size; 2) The null hypothesis is compared with several alternative hypotheses, making the test suitable in different experimental scenarios; 3) The test is constructed as a frequentist test, and defines significance with the conventional bound on Type I errors. We analyze the power of the test and find that it is higher than the power of other standard options, like the t-test (up to 25% higher) for a wide range of sample and effect sizes, and is at most 1% lower when the assumptions of the t-test are perfectly matched. We discuss and evaluate two variants of the test, that define different prior distributions over the parameters of the hypotheses.

## 1 Introduction

Most scientific papers revolve around a crucial point, where the reader needs to be convinced that the effect observed in the data or in numerical simulations is not the result of chance or noise, but is instead a robust outcome of the experiment. In the vast majority of cases, the task of persuading the reader is entrusted to a statistical significance test, based on the theory formulated in the 1920s and 1930s by Fisher [1], and Neyman and Pearson [2].

In many applications, and most notably in the life sciences, the experimental data often has characteristics that are ill-suited for standard frequentist tests: The sample size is often quite small ($N = 3, \ldots, 50$), as the cost and time required to collect data are high; moreover, it is often unclear whether the assumptions of the statistical tests are fulfilled (typically because of the reduced sample size). Under these conditions, standard tests tend to be very conservative, resulting in low statistical power and thus a loss of ability to resolve significant effects [3].

In this paper, we focus on the two sample problem, where the goal of the test is to decide whether samples from two independent populations share the same underlying distribution. We propose, as a general approach, to develop statistical tests based on Bayesian model selection. The Bayesian framework allows us to take into account the uncertainty in the parameters of the distributions arising from limited data, and to test the null hypothesis against several alternative hypotheses, thus constructing a test with good performance under different experimental conditions. We combine the ideas from model selection with standard methods for hypothesis testing, in order to be able to establish conventional false positive and false negative rates.

In the next two sections, we briefly summarize frequentist hypothesis testing and Bayesian data analysis. In Section 1.3 we describe a general hypothesis testing procedure based on Bayesian models selection. We then define a novel statistical test for two-sample, independent data (Sec. 2). In Section 3 we analyze the performance of the new test and compare its power against the standard t-test. The paper concludes with a discussion of related work.

## 1.1 Classical hypothesis testing

In the classical, frequentist approach to hypothesis testing, the experimenter chooses a statistical quantity, $t(\mathbf{y})$, that summarizes the observed data, $\mathbf{y}$, and is sensitive to the experimental effect to be tested. The researcher then formulates a *null hypothesis*, $H_0$, corresponding to a scenario with no experimental effect, and computes the distribution of $t$ under the null hypothesis.

Hypothesis testing is done, in the end, to reach a decision about the outcome of an experiment. This decision must be based on two factors: the value of the observed statistic, $t(\mathbf{y})$, and the cost of either rejecting the null hypothesis when there is no effect (*Type I error*), or failing to reject the null hypothesis in the presence of an effect (*Type II error*). The probability of a Type I error is also known as $\alpha$, or the *size* of the test, while the probability of a Type II error is $\beta$, or one minus the *power* of the test. Typically, in basic research areas like the neurosciences and cognitive sciences (in contrast, for example, to medical research), the cost of a Type I error far outweighs the consequences of committing an error of Type II. As a consequence, the rejection of the null hypothesis is based solely on keeping $\alpha$ below a critical threshold. By convention, the null hypothesis is rejected whenever the observed statistic, $t(\mathbf{y})$ lies above the 95 (or 99) percentile of the distribution of $t$ under $H_0$, thus fixing $\alpha$ to 0.05 (or 0.01). Given this decision rule, the performance of statistical tests can then be compared by their power, i.e., by their ability to detect experimental effects in relevant situations [4].

## 1.2 Bayesian data analysis and model selection

Bayesian statistics takes an alternative point of view regarding the validation of experimental data. According to Bayesians, every experiment begins with some prior knowledge or prior assumption about the data, in the form of a statistical model $H$ of data $\mathbf{y}$ with parameters $\theta$, and defined as the joint probability distribution $P(\mathbf{y}, \theta|H)$. After the observation of experimental data, $\mathbf{y}_{\text{exp}}$, the experimenter's beliefs are updated according to Bayes' theorem, which results in a posterior distribution over the parameters:

$$P(\theta|\mathbf{y} = \mathbf{y}_{\text{exp}}, H) = \frac{P(\mathbf{y} = \mathbf{y}_{\text{exp}}, \theta|H)}{P(\mathbf{y} = \mathbf{y}_{\text{exp}}|H)} \ . \quad (1)$$

In many treatments of Bayesian data analysis, the posterior distribution is the ultimate result of the experiment, and questions relative to hypothesis testing are answered by verifying if parameters values corresponding to a condition with no effect fall within a confidence interval in the posterior (e.g., if zero falls within a posterior confidence interval over the difference of the means of two populations) [5, 6].

Other authors have considered a second approach, based on Bayesian model selection, which is closer to the frequentist hypothesis testing approach [7, 8]. In model selection, the experimenter explicitly formulates a null hypothesis and an alternative hypothesis, $P(\mathbf{y}, \theta_0|H_0)$ and $P(\mathbf{y}, \theta_1|H_1)$. The two models are compared by computing the ratio of the probability of the data under the two models, known as *Bayes' factor*:

$$\frac{P(\mathbf{y}|H_1)}{P(\mathbf{y}|H_0)} = \frac{\int \mathrm{d}\theta_1 P(\mathbf{y}|\theta_1, H_1) P(\theta_1|H_1)}{\int \mathrm{d}\theta_0 P(\mathbf{y}|\theta_0, H_0) P(\theta_0|H_0)} \quad (2)$$

The alternative hypothesis is chosen over the null hypothesis whenever this ratio is sufficiently larger than one [7, 8]. The choice of a threshold is based on a general scale of what might be considered strong evidence for one hypothesis over the other, with a role similar to the 0.05 or 0.01 significance level in classical statistics. According to the scale proposed by H. Jeffrey [9], a ratio larger than 3 would correspond to "substantial evidence", larger than 10 to "strong evidence", and larger than 30 to "very strong evidence". It is important to note that the integral over parameters on the right side of Eq. 2 allows to compare models with a different number of parameters, due to its "automatic

Occam's razor" properties (Ref. 10, Ch. 28). Roughly speaking, the integral includes a penalty for more complex models, as the probability mass over parameters is spread over a larger volume. This is in contrast with likelihood ratio tests, that compute the ratio at the maximum likelihood parameters [4]. In that case, a higher number of parameters could lead to overfitting and thus to a higher frequency of Type I errors.

Bayesian data analysis is philosophically quite distant from the frequentist approach, and is based on the idea that researchers perform individual experiments, and not an infinite number over which the frequency of occurrence of a given event can be measured. As such, these methods do not guarantee a fixed Type I error. For example, it is not guaranteed that, for data drawn from a distribution with no effect, a no-effect value would be included 95% of the time in the 95% confidence interval of the posterior [7]. In fact, we will show in Section 3 that in some situations a Type I error of 0.05 would require the Bayes' factor to be considerably larger than Jeffrey's "strong evidence" bound, and that this threshold varies according to several factors, including the sample size. Another common critique of the Bayesian approach is that the final outcome of the test depends on the prior distribution over parameters ($P(\theta_0)$ and $P(\theta_1)$ in Eq. 2), and thus different experimenters may come to different conclusions based on their different subjective prior assumptions. Personally, we do not think that the choice of a prior is any different than the assumptions needed by standard frequentist tests, with the advantage of being explicitly stated and as such open to discussion. For example, a t-test implicitly assumes that the data is drawn from a normal distribution, and a prior over the mean of the data which is a delta function at the empirical mean. Nevertheless, because of these two issues Bayesian tests are rarely (if ever) used in practice for validating experimental results.

Bayesian models have, however, characteristics that are desirable for hypothesis testing. First of all, by encouraging the definition of explicit models for the data they offer the flexibility to define tests that are "customized" to the characteristics of the experiment. It would be very easy, for example, to define models for non-negative data, or with noise levels that depend on the value of the mean. Second, and most importantly, by considering a distribution over all the model's parameters they are able to take into account the uncertainty over parameters due to small sample sizes. This results in tests with higher statistical power for small sample sizes, as we will show in Section 3.

### 1.3 The model selection test

In this section, we propose a new approach that 1) combines the best qualities of the frequentist and Bayesian approach by using the Bayes' factor as the test statistics for a classical statistical test, and 2) is able to take into account several alternative hypotheses at once, and thus results in more flexible tests.

We define the test statistic to express the relative strength of belief that at least one of several alternative hypotheses, $H_i$, is more probable than the null hypothesis, $H_0$:

$$m(\mathbf{y}) = \max_i \left\{ \frac{P(\mathbf{y}|H_i)}{P(\mathbf{y}|H_0)} \right\} = \max_i \left\{ \frac{\int d\theta_i P(\mathbf{y}|\theta_i, H_i) P(\theta_i|H_i)}{\int d\theta_0 P(\mathbf{y}|\theta_0, H_0) P(\theta_0|H_0)} \right\} . \qquad (3)$$

High values of $m$ indicate that the observed data has higher probability under one of the alternative hypotheses than under the null hypothesis, and thus the null hypothesis is unlikely to be valid. Handling multiple alternative hypotheses is useful in increasing the flexibility of the test by achieving high power under several conditions that cannot be determined from the data. For example, in the two-samples test that we will derive in Section 2, we will consider two alternative hypotheses: that the data comes from two distributions with different means and equal ($H_1$) or different ($H_2$) variances. As discussed in the previous section, averaging over parameters ensures that all hypotheses are directly comparable, even if they vary in their complexity (i.e., in the number of parameters).

Redefining Bayesian model selection as a classical statistical test allows us to attain the same guarantees in terms of Type I and Type II errors than standard tests. Most importantly, the subjectivity deriving from the choice of the prior over parameters will only influence our conclusions as far as the power of the test is concerned, as the occurrence of a false positive is fixed by the Type I error. The subjective aspect of Bayesian statistics is thus folded into the decision threshold for the test statistic (see Table 2). This also has some consequences regarding the choice of the priors: In "pure" model selection one would generally define non-informative priors over parameters in order to make the comparison between models as fair as possible. In the model selection test, however, we are free to make more informative choices for the prior. The ideal choice is a compromise between capturing the a priori uncertainty about parameters due to observing a small number of samples, while restricting the prior space to exclude parameters that are very unlikely given the data in order to maximize the statistical power. In the results section, we will examine the effect of different choices for the prior.

## 2 Testing for differences in the mean

### 2.1 Definition of null- and alternative models

Based on these ideas, we proceed to develop a two-sample test, and compare its performance with standard alternatives like the t-test: Given data from two independent populations $\mathbf{y}^{(1)}$ and $\mathbf{y}^{(2)}$, we would like to determine whether the means of the underlying distributions are significantly different.

The proposed model selection test is based on three different hypotheses, the null hypothesis, $H_0$, and two alternative hypotheses, $H_1$ and $H_2$:

- $H_0$ is the null hypothesis, assuming no mean difference. The data is modelled as a single normal distribution with unknown mean, $\mu_0$, and variance, $\sigma_0^2$:

$$\mathbf{y}^{(1)}, \mathbf{y}^{(2)} \sim \text{Normal}(\mu_0, \sigma_0^2) \tag{4}$$

- $H_1$ assumes that the data was drawn from two normal distributions with unknown but different means, $\mu_1$ and $\mu_2$, and equal variances, $\sigma_{12}^2$:

$$\mathbf{y}^{(1)} \sim \text{Normal}(\mu_1, \sigma_{12}^2) \tag{5}$$
$$\mathbf{y}^{(2)} \sim \text{Normal}(\mu_2, \sigma_{12}^2) \tag{6}$$

- $H_2$ assumes that the data was drawn from two normal distributions with unknown but different means, $\mu_1$ and $\mu_2$, and different variances, $\sigma_1^2$ and $\sigma_2^2$:

$$\mathbf{y}^{(1)} \sim \text{Normal}(\mu_1, \sigma_1^2) \tag{7}$$
$$\mathbf{y}^{(2)} \sim \text{Normal}(\mu_2, \sigma_2^2) \tag{8}$$

Our choice for the prior distributions over the model parameters, $\mu_i$ and $\sigma_i$, was informed by two principles: first, the uncertainty over the mean of the distributions should be inversely related to the number of data points. We thus define the prior over the means to be a normal distribution with a standard deviation that scales as the standard error of the mean (SEM), i.e., as $1/\sqrt{N}$. This implies that the uncertainty over the mean under $H_0$ will be smaller than that of the means under $H_1$ and $H_2$, since $H_0$ considers the data points from both populations. Second, we choose a broad, uniform distribution as the prior over the standard deviation of the data.

In order to simplify the notation, without loss of generality, we assume in the following that the observed data $\mathbf{y} = \{\mathbf{y}^{(1)}, \mathbf{y}^{(2)}\}$ has zero mean and an empirical variance of 1, which can always be achieved by shifting and rescaling the data. The priors over parameters are defined as follows:

- Under the null hypothesis, $H_0$ (Eq. 4), the prior over the mean of this distribution, $\mu_0$, is a normal distribution centered at the empirical mean, $\overline{y}$, with standard deviation equal to $1/\sqrt{N}$. The prior concentrates around the empirical mean as the number of total samples increases, reflecting a decrease in the uncertainty over that parameter. For the standard deviation, $\sigma_0$, we choose an interval bounded below by zero (since $\sigma_0$ is a positive quantity), and above by 3 times the empirical standard deviation of the data (which is 1 due to normalization). The interval we chose is quite conservative, as the mean would need to be far outside the observed range for the real standard deviation to reach that value. In summary,

$$\mu_0 | H_0 \sim \text{Normal}\left(\overline{y}, \frac{1}{N}\right) \tag{9}$$
$$\sigma_0 | H_0 \sim \text{Uniform}(\epsilon, 3) \tag{10}$$

where $\epsilon = 10^{-3}$ is a value close to zero to avoid numerical errors.

- For the first alternative hypothesis, $H_1$ (Eqs. 5-6), the priors over the parameters are defined as for $H_0$, with the exception of the range over $\sigma_{12}$, which was decreased to 1 in order to reflect the fact that the data from the two individual populations must have a smaller standard deviation than the combined data:

$$\mu_1|H_1 \sim \text{Normal}\left(\overline{\mathbf{y}^{(1)}}, \frac{1}{N_1}\right) \quad (11)$$

$$\mu_2|H_1 \sim \text{Normal}\left(\overline{\mathbf{y}^{(2)}}, \frac{1}{N_2}\right) \quad (12)$$

$$\sigma_{12}|H_1 \sim \text{Uniform}(\epsilon, 1) \quad (13)$$

- Finally, priors for the second alternative hypothesis, $H_2$, were defined similarly as for $H_1$:

$$\mu_1|H_2 \sim \text{Normal}\left(\overline{\mathbf{y}^{(1)}}, \frac{1}{N_1}\right) \quad (14)$$

$$\mu_2|H_2 \sim \text{Normal}\left(\overline{\mathbf{y}^{(2)}}, \frac{1}{N_2}\right) \quad (15)$$

$$\sigma_1|H_2 \sim \text{Uniform}(\epsilon, 1) \quad (16)$$

$$\sigma_2|H_2 \sim \text{Uniform}(\epsilon, 1) \quad (17)$$

In Section 3 we will consider two alternative choices for the prior distributions: a more informative one that makes use of empirical statistics of the data, and a less informative one that assigns the same prior to all parameters, and analyze how the power of the test changes when the prior is excessively informative or not enough so.

## 2.2 Computing the Bayes' factor

In order to compute the test statistics, $m(\mathbf{y})$, we compute the integrals over parameters (Eq. 3) by Monte Carlo integration [11]:

$$P(\mathbf{y}|H_i) = \int d\theta_i \, P(\mathbf{y}|\theta_i, H_i) \, P(\theta_i|H_i) \simeq \frac{1}{N_{\text{samples}}} \sum_{\theta_i^* \sim P(\theta_i|H_i)} P(\mathbf{y}|\theta_i^*, H_i) \;, \quad (18)$$

for $i \in \{0, 1, 2\}$. Since the number of parameters and their range is small, the integral can be computed quite efficiently. In our simulations, the integral is computed with 1500 samples, ensuring an approximation of the final value with only a small variance.

Alternatively, one can compute the integral using the posterior distribution over parameters [11]:

$$\frac{1}{P(\mathbf{y}|H_i)} = \int d\theta_i \frac{P(\theta_i|\mathbf{y}, H_i)}{P(\mathbf{y}|\theta_i, H_i)} \simeq \frac{1}{N_{\text{samples}}} \sum_{\theta_i^* \sim P(\theta_i|\mathbf{y}, H_i)} \frac{1}{P(\mathbf{y}|\theta_i^*, H_i)} \;. \quad (19)$$

In theory, this alternative should be more efficient because the posterior distribution is concentrated in the space of relevant parameters. However, the reduction in variance of the estimation is compensated by the fact that the parameters become correlated when conditioned on the data. Thus, unless there is an analytical expression for the posterior, sampling from it requires a larger number of samples (for the burn-in phase and thinning). We found that, for our test, this second alternative required substantially more time.

## 3 Results

In order to analyze the Type I error of the test, we first need to derive the distribution of the summary statistics, $m$. This depends on the number of samples in the two populations, $N_1$ and $N_2$. Thus, we computed tables for varying $N_1$ and $N_2$ by repeatedly drawing random data from the null hypothesis model, $H_0$, and computing $m$ according to Eq. 3 (100,000 runs per table).

Using the tables we computed the Type I error that one would commit using Jeffrey's scale as a threshold for the Bayes' factor to decide whether to reject the null hypothesis. As shown in Table 1, accepting evidence for the alternative hypothesis based on "very strong" evidence on Jeffrey's scale ($m > 30$) always results in Type I error smaller than 0.05. For a wide range of parameter settings, however, the limit for "substantial" ($m > 3$) and even "strong" evidence ($m > 10$) would not reach the conventional 0.05 significance limit. Conversely, we can look at the threshold on $m$ necessary to ensure a Type I error of 0.05. The bottom line of Table 1 shows that the threshold on the Bayes' factor decreases with the number of samples to approximately 8 for $N_{1/2} = 50$ and 100. As a consequence, using a high Bayes' factor as a criterion for the rejection of the null hypothesis does not guarantee a low Type I error. Furthermore, whether a high Bayes' factor implies low Type I error depends on the data (in this case, on the number of samples in the data), and, more in general, on the prior distribution over parameters (Table 2).

| | | $N_{1/2}$ | | | | | |
|---|---|---|---|---|---|---|---|
| | | 3 | 4 | 5 | 10 | 50 | 100 |
| evidence | substantial, $m > 3$ | 0.22 | 0.20 | 0.18 | 0.16 | 0.15 | 0.15 |
| | strong, $m > 10$ | 0.09 | 0.07 | 0.07 | 0.05 | **0.04** | **0.04** |
| | very strong, $m > 30$ | **0.04** | **0.03** | **0.02** | **0.02** | **0.01** | **0.01** |

| threshold for $\alpha = 0.05$ | 19.8 | 15.7 | 13.7 | 10.2 | 7.8 | 7.7 |
|---|---|---|---|---|---|---|

**Table 1: Type I error associated with Jeffrey's evidence scale.** Top three rows: Type I error committed by rejecting the null hypothesis when the Bayes' factor in Eq. 3 is larger than the thresholds of Jeffrey's evidence scale. Bold entries indicate that the Type I error is below the conventional significance limit of 0.05 . Bottom row: minimum threshold necessary to achieve a Type I error of 0.05 .

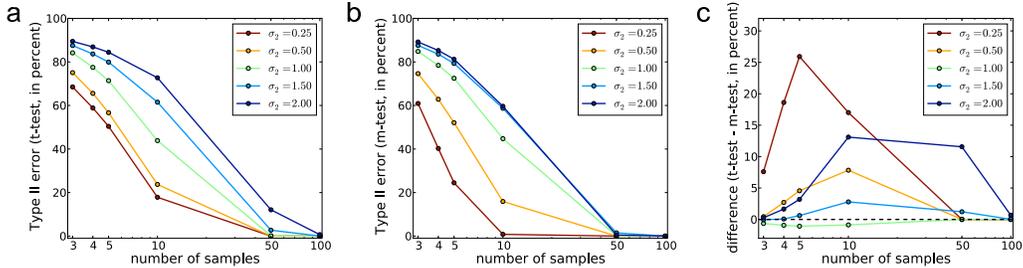

**Figure 1: Comparison of Type II error in t-test and m-test.** **a,b**, Type II error for t-test (**a**) and model selection test (m-test, **b**) as a function of the number of samples in each population (x-axis) and of the standard deviation of the second population (colors). **c**, Difference in Type II error between t-test and m-test. Positive numbers indicate lower error (greater statistical power) for the m-test.

In frequentist statistical tests, the critical thresholds on the statistic is set such that the probability of a Type I error is equal to a certain acceptable level. Even though the frequency of false positives is fixed in this way, statistical tests still differ in their Type II error, with better tests having a lower Type II error for a fixed Type I error. We computed the Type II error by sampling 3 to 100 data points from two normal distributions, one with zero mean and unit variance, and the second with mean 1 and standard deviation between 0.25 and 2. Type I error was kept fixed at 5% by rejecting the null hypothesis whenever $m$ was larger than the threshold in Table 1 (bottom row). Figure 1 shows the Type II error over 50,000 runs for the t-test (Fig. 1a) and the model selection test (Fig. 1b). For both tests, the Type II error decreases with the number of samples and with decreasing standard deviation of the second population, as $\mathbf{y}^{(1)}$ and $\mathbf{y}^{(2)}$ become more separated. The Type II error of the model selection test shows a remarkable improvement of up to 25% over the t-test for the vast majority of cases (Fig. 1c). The single exception is when the assumptions of the t-test are exactly matched, with the two population distributions having the same variance, in which case there is a small increase in Type II error of at most 1% (Fig. 1c, green line). The model selection test is thus more flexible and powerful than the t-test in a wide range of situations.

One might argue that the reason the t-test performs worse with respect to our test is that it assumes that the two populations have the same variance. There exist standard alternatives to the t-test for the case where the variances of the two populations are different, most notably Welch's test [12], and the non-parametric Mann-Whitney U test (Ref. 13; see Ref. 14 for a review of the attempts to solve the two-sample problem). However, previous studies have shown that the power of these alternative tests is not greater than that of the t-test, even when the variances are different [3]. In particular, Figure 2b shows that Welch's test has a slightly higher Type II error than the t-test under the conditions of our experiments, in agreement with the findings in [3]. Ref. 3 also considers a non-parametric test known as Lapage's test which has lower Type II error than the t-test when the variances are very different, but substantially higher when they are similar to each other (e.g., 8% higher when $N_{1/2} = 25$, the difference between means is one, and both populations have variance of 1).

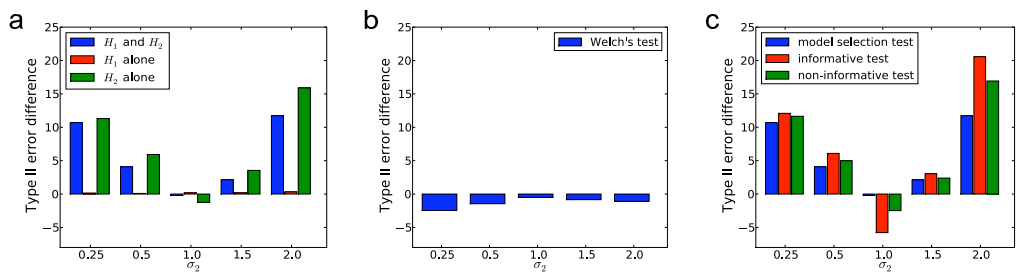

**Figure 2: Comparison of t-test Type II error with other tests.** Average difference in Type II error between t-test and other tests. Positive number indicate an improvement in Type II with respect to the t-test. The average is taken by numerical integration (trapezoidal rule) over $N_{1/2} = 3, \ldots, 50$ of the difference curves (cf. Fig. 1c). **a**, Comparison with the m-test defined with both alternative hypothesis (blue), only the same-variance hypothesis (red), and only the different-variances hypothesis (red). **b** Comparison with Welch's t-test. In agreement with [3], Welch's test has a slightly higher Type II error under these conditions. **c**, Comparison with three different priors for the parameters of the m-test. blue: The original test; red: An informative test that makes extensive use of the data to define the prior distributions; green: A non-informative test with priors that are almost independent from the data.

An important practical issue is deciding when the same variance assumption is violated, in order to apply appropriate tests. A common practice, implemented in statistical software like SAS and SPSS, consists of preceding the two-sample test with an F-test on the variances of the population. If the two variances are significantly different, another test (e.g., a Welch test) is used instead of the t-test. However, the sequential test composed of the F-test followed by a two-sample test actually increases the Type I error, and leads ultimately to a loss of power [14, 15]. Thanks to the combination of multiple alternative hypotheses, the model selection test has the flexibility to perform well on a wide range of sample sizes and differences in variance. Figure 2 shows the average Type II error for different variance ratios of the complete test, compared with the Type II error of a test that only considers one of the two alternative hypotheses. As may have been expected, $H_1$ alone performs better when the two variances are equal, while $H_2$ alone performs better when they are not. The combined test has properties that are intermediate between the two, and very close to the best of the two for each case. As a consequence of the increased flexibility, the experimenter does not have to worry about violations of the assumptions regarding similar or different variances, and can directly apply the test. Taking into account additional alternative hypotheses might provide a low Type II error in a wider range of situations, as for example departures from normality.

Finally, we consider the effect of using two alternative choices for the priors over the parameters of the hypotheses (Eqs. 9–17). In the first test, the priors over the means are defined as normal distributions centered at the empirical mean, with standard deviation given by the SEM of the data. The prior over the standard deviation is defined as a uniform interval between zero and the empirical standard deviation one would obtain if the actual mean was displaced from the empirical mean by three times the SEM (see Supplementary Material for a detailed definition of the priors). Thus, these priors are very informative, as they derive estimates about the uncertainty in the real underlying parameters directly from the data. In the second test, the priors over the parameters are the same for all models, and depend very little on the observed data. The prior over the mean is a normal distribution centered at the empirical mean, with unit standard deviation of 1. The prior over the standard deviation is a uniform distribution between 0 and 3. These priors are the least informative of the three tests. Figure 2c shows the average improvement in Type II error over the t-test for the three tests (the complete Type II error curves are shown in Supplementary Figure 1). In both cases, the alternative priors show an improvement in power for the cases where the variances of the two populations are most dissimilar. However, this is compensated by a loss in power when the two variances are the same. These results confirm that the choice of the priors over the parameters should not be overly informative, as the risk of overfitting the data leads to high thresholds on $m$ to avoid an excessive false positive rate. Nor should it be too generic, which might lead to low power in some situations. This is also reflected in the thresholds on $m$ needed to achieve $\alpha \leq 0.05$ (Table 2): the critical values for the informative prior are very high, reflecting a high frequency of false positives due to overfitting; the thresholds for the non-informative test are instead quite low because of their looseness.

|  | $N_{1/2}$ | | | | | |
|---|---|---|---|---|---|---|
|  | 3 | 4 | 5 | 10 | 50 | 100 |
| model selection test | 19.8 | 15.7 | 13.7 | 10.2 | 7.8 | 7.7 |
| informative test | 66.2 | 32.4 | 20.8 | 10.8 | 4.2 | 2.9 |
| non-informative test | 7.7 | 6.9 | 6.0 | 4.0 | 1.6 | 1.3 |

Table 2: **Thresholds for statistical significance in variants of the m-test.** Minimum threshold on $m$ necessary to achieve a 0.05 Type I error for the model selection test and the two variants discussed in the text.

## 4 Discussion

In this paper, we introduced a novel two-sample test combining Bayesian model selection with the standard frequentist hypothesis testing framework, and showed an increase in the power of the test with respect to classical two-sample tests. The model selection test is very flexible, as it allows one to take into account multiple alternative hypotheses, each indicating a possible experimental scenario. In the two-sample test presented here, this allowed the test to perform well for both similar and very different variances in the two populations.

One of the reasons we expect tests based on model selection to perform better than their standard counterpart is because they take into account the uncertainty over the parameters. The advantage of using the model selection test will thus be larger for small number of samples. In fact, we see from our results that the difference is greater for sample sizes between 3 and 50. This is the regime of a very large body research in the life sciences, including the neurosciences and cognitive sciences, where the cost of collecting data is often very high (e.g., studies of microarray data typically have $N_{1/2} = 1$ to 3 [16]; for psychophysical experiments, $N = 20$ is a typical sample size). In such applications, an increase of in statistical power is crucial.

Two previous studies [17, 16] made use of prior information, coming from estimates of the variance from related experiments, to improve on the t-test when such information is available. By analytical integration of the prior distributions over mean and standard deviation they derived an expression for the degrees of freedom of the t-statistic. As a result, they were able to obtain an improvement in power for very small sample sizes. Similarly, one might choose prior distributions over the parameters in the model selection test such that it is possible to carry out the integrals in Eq. 3 analytically. Borgwart and Ghahramani [8] showed that this is possible if the likelihood terms, $P(\mathbf{y}|\theta_i, H_i)$, are in the exponential family, and one uses conjugate priors. However, it is at present unclear whether this class of priors is not too diffuse to result in a test with high power. For example, a inverse Gamma prior on the variance would put some probability mass on very large variances that are unlikely to occur, lowering the marginal probability of the data given the model. Using Monte Carlo integration, as in this paper, allows us to define arbitrary and potentially more complex models.

We think that the model selection approach presented here is simple and flexible enough to be applicable in a wide range of practical situations. This method encourages researchers to make use of customized statistical tests that exploit all of the information at hand for their data, resulting in more powerful tests.

# Supplementary material
# A frequentist two-sample test
# based on Bayesian model selection

**Pietro Berkes and József Fiser**
Volen Center for Complex Systems
Brandeis University, Waltham, MA 02454

## 1 Definition of the alternative model selection tests

We report here the detailed definition of the priors over parameters for the two alternative model selection tests described in Section 3 of the main text.

In the first "informative" test, the priors over the means are defined as normal distributions centered at the empirical mean, with standard deviation given by the standard error of the mean (SEM) of the data. The prior over the standard deviation is defined as a uniform interval between zero and the empirical standard deviation one would obtain if the actual mean was displaced from the empirical mean by 3 times the SEM, i.e.,

$$\sigma_{3\text{SEM}}(\mathbf{y}) = \sqrt{\frac{1}{N-1} \sum_i (\mathbf{y}_i - (\overline{\mathbf{y}} + 3 \cdot \text{SEM}(\mathbf{y})))^2} \tag{1}$$

In summary:

$$\mu_0 | H_0 \sim \text{Normal}(\overline{\mathbf{y}}, \text{SEM}(\mathbf{y})^2) \tag{2}$$
$$\sigma_0 | H_0 \sim \text{Uniform}(\epsilon, \sigma_{3\text{SEM}}(\mathbf{y})) \tag{3}$$

$$\mu_1 | H_1 \sim \text{Normal}\left(\overline{\mathbf{y}^{(1)}}, \text{SEM}(\mathbf{y}^{(1)})^2\right) \tag{4}$$
$$\mu_2 | H_1 \sim \text{Normal}\left(\overline{\mathbf{y}^{(2)}}, \text{SEM}(\mathbf{y}^{(2)})^2\right) \tag{5}$$
$$\sigma_{12} | H_1 \sim \text{Uniform}(\epsilon, \max(\sigma_{3\text{SEM}}(\mathbf{y}^{(1)}), \sigma_{3\text{SEM}}(\mathbf{y}^{(2)}))) \tag{6}$$

$$\mu_1 | H_2 \sim \text{Normal}\left(\overline{\mathbf{y}^{(1)}}, \text{SEM}(\mathbf{y}^{(1)})^2\right) \tag{7}$$
$$\mu_2 | H_2 \sim \text{Normal}\left(\overline{\mathbf{y}^{(2)}}, \text{SEM}(\mathbf{y}^{(2)})^2\right) \tag{8}$$
$$\sigma_1 | H_2 \sim \text{Uniform}(\epsilon, \sigma_{3\text{SEM}}(\mathbf{y}^{(1)})) \tag{9}$$
$$\sigma_2 | H_2 \sim \text{Uniform}(\epsilon, \sigma_{3\text{SEM}}(\mathbf{y}^{(2)})) \tag{10}$$

In the second "non-informative" test, the prior over the mean is always a normal distribution centered at the empirical mean, with unit standard deviation of 1. The prior over the standard deviation is always a uniform distribution between 0 and 3:

$$\mu_0|H_0 \sim \text{Normal}(\overline{\mathbf{y}}, 1) \tag{11}$$
$$\sigma_0|H_0 \sim \text{Uniform}(\epsilon, 3) \tag{12}$$

$$\mu_1|H_1 \sim \text{Normal}\left(\overline{\mathbf{y}^{(1)}}, 1\right) \tag{13}$$
$$\mu_2|H_1 \sim \text{Normal}\left(\overline{\mathbf{y}^{(2)}}, 1\right) \tag{14}$$
$$\sigma_{12}|H_1 \sim \text{Uniform}(\epsilon, 3) \tag{15}$$

$$\mu_1|H_2 \sim \text{Normal}\left(\overline{\mathbf{y}^{(1)}}, 1\right) \tag{16}$$
$$\mu_2|H_2 \sim \text{Normal}\left(\overline{\mathbf{y}^{(2)}}, 1\right) \tag{17}$$
$$\sigma_1|H_2 \sim \text{Uniform}(\epsilon, 3) \tag{18}$$
$$\sigma_2|H_2 \sim \text{Uniform}(\epsilon, 3) \tag{19}$$

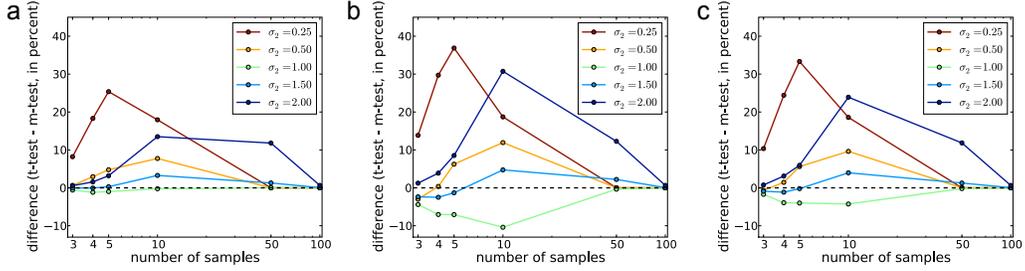

**Figure 1: Comparison of t-test Type II error with other tests.** Difference in Type II error between t-test and the three variant of the model selection test described at the end of Section 3: **a**, the main test, **b**, the test with informative priors , and **c**, the test with non-informative priors. Positive numbers indicate lower error (greater statistical power) than the t-test.